\documentclass[aps,prl,10pt, tightenlines, twoside, secnumarabic, 
superscriptaddress,showpacs, preprintnumbers, showpacs, nofootinbib, 
twocolumn]{revtex4-1}

\pdfoutput=1
\usepackage[english]{babel}
\usepackage{amsmath,amssymb,amsfonts, bm,bbm,slashed}
\usepackage{graphicx}
\usepackage[sort&compress]{natbib}
\usepackage{xcolor}
\usepackage[normalem]{ulem}
\usepackage{hyperref}
\usepackage{cleveref}
\definecolor{red}{rgb}{1.0, 0, 0}
\usepackage{hyperref}
\usepackage{enumerate}
\usepackage{epsfig, subfigure}
\usepackage{setspace}
\usepackage{booktabs, tabularx}
\usepackage{units}

\hypersetup{
    pdfnewwindow=true,      
    colorlinks=true,       
    linkcolor=blue,          
    citecolor=blue,        
    filecolor=blue,      
    urlcolor=blue        
}

\allowdisplaybreaks

\bibpunct{[}{]}{,}{n}{}{,}

\newcommand{\ev}[1]{\ensuremath{\left\langle #1 %
                     \right\rangle}} 
\newcommand{\tr}{\text{tr}}

\renewcommand{\vec}[1]{{\mathbf{#1}}}

\begin{document}

\title{A m\'enage \`a trois of eV-scale sterile neutrinos, cosmology, and structure formation}

\author{Basudeb Dasgupta}
\email{bdasgupta@ictp.it}
\affiliation{International Centre for Theoretical Physics, Strada Costiera 11, 34014 Trieste, Italy.}

\author{Joachim Kopp}
\email{jkopp@mpi-hd.mpg.de}
\affiliation{Max Planck Institut f\"ur Kernphysik, Saupfercheckweg 1, 69117 Heidelberg, Germany.}

\date{December 11, 2013}

\begin{abstract}
We show that sterile neutrinos with masses $\gtrsim 1$~eV, as motivated by
several short-baseline oscillation anomalies, can be consistent with
cosmological constraints if they are charged under a hidden sector force
mediated by a light boson.  In this case, sterile neutrinos experience a large
thermal potential that suppresses mixing between active and sterile
neutrinos in the early Universe, even if vacuum mixing angles are large.  Thus,
the abundance of sterile neutrinos in the Universe remains very small, and their
impact on Big Bang Nucleosynthesis, Cosmic Microwave Background, and 
large-scale structure formation is negligible.  It is conceivable that the new 
gauge force also couples to dark matter, possibly ameliorating some of the 
small-scale structure problems associated with cold dark matter.
\end{abstract}

\begin{flushright}
\end{flushright}

\maketitle

\section{Introduction}
\label{sec:intro}
Several anomalies in short baseline neutrino oscillation experiments have
spurred interest in models with more than three neutrino species. In
particular, the excesses of electron neutrino events in LSND~\cite{Aguilar:2001ty} and
MiniBooNE~\cite{AguilarArevalo:2012va}, as well as the unexpected
electron antineutrino disappearance at short baselines~\cite{Mueller:2011nm,
Mention:2011rk, Hayes:2013wra, Acero:2007su}, could be explained in models with
extra ``sterile'' neutrinos, i.e.\ light ($m\sim 1$~eV) new fermions that are
uncharged under the Standard Model (SM) gauge group and mix with the three
known neutrino species.  On the other hand, a number of other neutrino
oscillation experiments that did not observe any anomalous signals put such
models under pressure~\cite{Kopp:2011qd, Kopp:2013vaa, Conrad:2012qt, Kristiansen:2013mza, Giunti:2013aea},
and a vigorous experimental program is currently underway to resolve the
tension by either confirming the anomalies, or by providing a definitive null
result.

It is often argued that the tightest constraints on sterile neutrino models
come from cosmology. Indeed, the simplest models --- with just one
or several sterile neutrinos, but no other new particles --- are disfavored by
the Big Bang Nucleosynthesis (BBN) and Planck measurements of $N_\text{eff}$, 
the number of relativistic particle
species in the early Universe~\cite{Steigman:2012ve, Ade:2013lta}. 
For sterile neutrino masses of $\sim 1$~eV, or larger, even tighter
constraints are obtained from large-scale structure
formation~\cite{Hamann:2011ge}, where the presence of extra neutrino species
would lead to a wash-out of structure due to efficient energy
transport by neutrinos.

In this paper, we show that these constraints are evaded if sterile neutrinos
have hidden interactions mediated, for instance, by a new gauge boson $A'$,
often called a dark photon, with a mass $M\lesssim \text{MeV}$.
As discussed below, gauge forces of this type are also interesting in dark matter (DM) physics,
and are probed in many cosmological and astrophysical searches~\cite{Jaeckel:2010ni}.
We will show that at non-zero temperature the sterile neutrinos
feel a Mikheyev-Smirnov-Wolfenstein (MSW) potential that suppresses
mixing between active and sterile neutrinos in the early Universe, thus preventing
sterile neutrino production in the early Universe. We will discuss constraints
on this scenario from cosmology and particle physics. In the last part of the
paper we will also discuss the possibility that $A'$ couples also to the
DM ($\chi$) in the Universe, possibly easing the disagreement between small-scale 
structure observations and cold DM simulations.

\section{Hidden Sterile Neutrinos}
\label{sec:mechanism}
We assume the Standard Model (SM) is augmented by one extra species of light
($\sim \text{eV}$) neutrinos $\nu_s$, which do not couple to the SM gauge bosons
but are charged under a new $U(1)_\chi$ gauge symmetry. We assume that $\nu_s$
have relatively large ($\sim 10\%$) vacuum mixing with the active neutrinos and
is thus capable of explaining the short baseline oscillation anomalies.

The sterile sector is expected to be coupled to the SM sector through high-scale
interactions, and the two sectors decouple at
temperatures $\gtrsim \text{TeV}$. Our results remain qualitatively
correct even for decoupling temperatures as low as 1~GeV, i.e.,\ just above the
QCD phase transition. After decoupling, the temperature $T_s$ of the sterile sector
continues to drop as $T_s \sim 1 / a$ ($a$ being
the scale factor of the Universe), while the temperature in the visible sector,
$T_\gamma$, drops more slowly because of the entropy generated when heavier
degrees of freedom (unstable hadrons, positrons, etc.) become inaccessible and
annihilate or decay away. By the BBN epoch, the number of effective 
degrees of freedom of the visible sector, $g_*$, decreases from $\simeq106.7$ to $\simeq10.75$. 
Taking the sterile sector temperature as 
$T_s  =(g_{*,{\rm T}_\gamma}/g_{*,{\rm TeV}})^{{1}/{3}} T_\gamma$, 
the additional effective number of fully-thermalized neutrinos at BBN, for a single
left-handed sterile neutrino (and its right-handed antineutrino) and a relativistic $A'$, is
\begin{align}
\Delta N_\nu &\equiv\frac{\rho_{\nu_s}+\rho_{A'}}{\rho_{\nu}}=\frac{(g_{\nu_s} + g_{A'})\,T_s^4}{g_\nu\,T_\nu^4}\\
             &=\frac{\left(\tfrac{7}{8}\times2 + 3\right) \times\left(\tfrac{10.75}{106.7}\right)^{\tfrac{4}{3}}}
             {\left(\tfrac{7}{8}\times2\right)\times\left(\tfrac{4}{11}\right)^{\tfrac{4}{3}}}\simeq0.5\,,
\end{align}
which is easily consistent with the bound from BBN, viz., $\Delta N_\nu=0.66^{+0.47}_{-0.45}$~\cite{Steigman:2012ve}. 
Up to 3 generations of sterile neutrinos could be accommodated within $\simeq 1\sigma$. Note that we have conservatively 
taken $T_\nu$ at the end of BBN.

At lower temperatures, $T_s\lesssim 0.1\,$MeV, $A'$ becomes nonrelativistic, 
and decays to sterile neutrinos, heating them up by a factor of 
$\simeq1.4$. However, these neutrinos with masses $m\gtrsim1\,$eV, are 
nonrelativistic by the epoch of matter-radiation equality 
($T_\gamma\simeq 0.7\,$eV) and recombination ($T_\gamma\simeq0.3\,$eV).
Thus the impact of thermal abundances of $A'$ and $\nu_s$ on the CMB and 
structure formation is negligible. See also \cite{Foot:1995bm, Barger:2003rt, Ho:2012br} 
for alternate approaches. We will now show that oscillations of 
active neutrinos into sterile neutrinos, which are normally expected to bring the two 
sectors into equilibrium again, are also strongly suppressed due to 
``matter'' effects.
 
The basic idea underlying our proposal is similar to the high-temperature counterpart 
of the MSW effect. Let us recall that at high temperatures, i.e., in the 
early Universe, an active neutrino with energy $E$ experiences a potential  
$V_{\rm MSW}\propto G_F^2 E T_\gamma^4$ due to their own 
\emph{energy density}~\cite{Notzold:1987ik}. This  
is not zero even in a CP symmetric Universe. A similar, but much larger, potential 
can be generated at high-temperature for sterile neutrinos, 
\emph{if} they couple to a light hidden gauge boson $A'$. There are two types of processes that can contribute 
to this potential --- the sterile neutrino can forward-scatter off an $A'$ in the 
medium, or off a fermion $f$ that couples to $A'$. 

These interactions of the sterile neutrino with the medium 
modify its dispersion relation through a potential $V_{\rm eff}$:
\begin{align}
E=|\vec{k}|+\frac{m^2}{2E}+V_{\rm eff}\,,
\end{align}
where $E$ and $|\vec{k}|$ are the energy and momentum of the sterile neutrino. 

\begin{figure}[!t]
  \begin{center}
    \vspace{-0.3cm}
    \includegraphics[width=0.95\columnwidth]{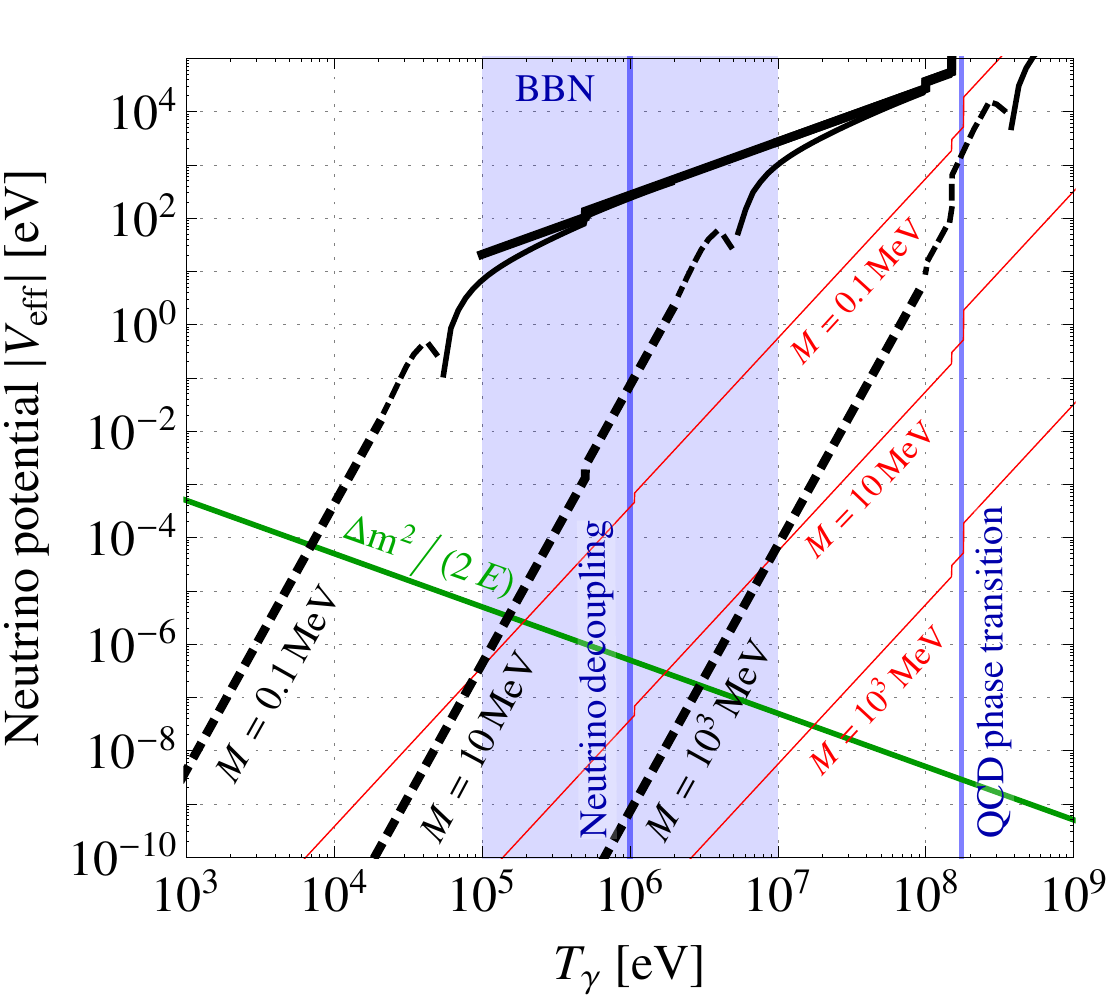}
  \end{center}
  \vspace{-0.5cm}
  \caption{Comparison of the effective matter potential $V_{\rm eff}$ for 
    sterile neutrinos (black curves) to the active--sterile oscillation frequency
    $\Delta m^2 / (2 E)$ (green line) at $E\simeq T_\gamma$ and $\Delta m^2 =
    1$~eV$^2$. As long as $|V_\text{eff}|\gg\Delta m^2 / (2 E)$,
    oscillations are suppressed. Different black curves show $|V_\text{eff}|$ for
    different values of the gauge boson mass $M$, with solid lines corresponding
    to $V_\text{eff} > 0$ and dashed lines indicating $V_\text{eff} < 0$. Thin
    (Thick) lines show exact numerical (approximate analytical) results. The
    hidden sector fine-structure constant is taken as $\alpha_\chi\equiv
    e_\chi^2/(4\pi)=10^{-2}/(4\pi)$. Red lines show the contribution to
    $V_\text{eff}$ from an asymmetric DM  particle with $m_\chi = 1$~GeV. The QCD
    phase transition and active neutrino decoupling epochs are annotated. The
    small kinks in the curves are due to changes in $g_*$, the effective number
    of degrees of freedom in the Universe.}
  \label{fig:Vplot}
\vspace{-0.7cm}  
\end{figure}

We calculated $V_{\rm eff}$ using the real time formalism in thermal field theory 
(see Appendix A). Physically, this potential is the 
correction to the sterile neutrino self-energy.
In the low-temperature limit, i.e., $T_s, E \ll M$, we find
$V_\text{eff} \simeq -{28\pi^3 \alpha_\chi E T_s^4}/{(45 M^4)} \,,$
similar to the potential for active neutrinos~\cite{Notzold:1987ik}, 
with $\alpha_\chi\equiv e_\chi^2/(4\pi)$ being the $U(1)_\chi$ fine-structure 
constant. In the high-temperature limit, $T_s, E\gg M$, we find
$V_\text{eff} \simeq +{\pi\alpha_\chi T_s^2}/{(2 E)}\,,$
similar to the result for hot QED~\cite{Weldon:1982bn}. We have assumed that 
there is no asymmetry in $\nu_s$, which may be interesting to 
consider~\cite{Foot:1995bm, Saviano:2013ktj}.
These analytical results are plotted in Fig.\,\ref{fig:Vplot} (thick black lines).
For comparison, we also calculated the potential numerically (thin black lines), and found 
excellent consistency with the analytical approximations in their region of validity. 
The potential is small only in a very small range of temperatures $T_s\approx M$, where 
the potential changes sign and goes through 
zero. Note that the potential is always smaller that $|\vec{k}|$ and 
vanishes at zero temperature.

In the presence of a potential, it is well-known that neutrino mixing angles 
are modified. In the two-flavor approximation, the
effective mixing angle $\theta_m$ in matter is given by~\cite{Akhmedov:1999uz}
\begin{align}
  \sin^2 2\theta_m
    = \frac{\sin^2 2\theta_0}
           {\left(\cos 2\theta_0 + \tfrac{2 E}{\Delta m^2} V_\text{eff}\right)^2 + \sin^2 2\theta_0} \,,
\end{align}
where $\theta_0$ is the vacuum mixing angle, and $\Delta m^2 = m_s^2 - m_a^2$ is the difference between
the squares of the mostly sterile mass eigenstate $m_s$ and the active neutrino
mass scale $m_a$. If the potential is much larger than the vacuum oscillation 
frequency, i.e.,
\begin{align}
  |V_\text{eff}| \gg \bigg| \frac{\Delta m^2}{2 E} \bigg| \,,
  \label{eq:Veff-condition}
\end{align}
then $\theta_m$ will be tiny, and oscillations of active neutrinos into sterile ones
are suppressed.

This is confirmed by Fig.~\ref{fig:Vplot}, which summarizes our main results.
For a typical neutrino energy $E \sim T_\gamma$ and $M \lesssim 10$~MeV, we
see that condition \eqref{eq:Veff-condition} is well-satisfied down to temperatures
$T_\gamma \lesssim 1$~MeV, i.e.,\ until after the time of neutrino decoupling, when
their thermal production becomes impossible.
Thus $\theta_{m}$ is suppressed and 
sterile neutrinos are not produced in significant numbers. 
There is also non-forward scattering of sterile neutrinos mediated by the hidden gauge boson, as well as the usual MSW potential for active neutrinos, which further suppress oscillations. A full numerical calculation using quantum kinetic equations~\cite{Hannestad:2013ana} is consistent with our simple estimate using condition \eqref{eq:Veff-condition}. Oscillations after 
decoupling reduces a small fraction, $\sin^2 2\theta_m\lesssim 0.1$, of the active neutrinos to 
steriles (which are nonrelativistic below $1\,$eV), consistent with 
$N_{\rm eff}=3.30^{+0.54}_{-0.51}$ (95\% limits) from cosmological data~\cite{Ade:2013lta}. 
Note that in Fig.~\ref{fig:Vplot}, we have conservatively taken sterile neutrino decoupling to 
occur at the same temperature, $T_\gamma \simeq 1$~MeV, as the decoupling of active 
neutrinos. In reality, sterile neutrino production ceases when $\Gamma_s \sim \sin^2 \theta_s G_F^2 T_\gamma^5$
drops below the Hubble expansion rate $H \propto T_\gamma^2$, which happened at 
temperatures around $1\ \text{MeV} / (\sin^2\theta)^{1/3}$.

Even for $M$ slightly larger than 1~MeV,
sterile neutrino production remains suppressed until the BBN epoch, but it
is interesting that in this case $V_\text{eff}$ crosses zero
while neutrinos are still in thermal equilibrium. 
This implies that there is a brief
time-period during which sterile neutrinos could be produced efficiently. 
However, as long as its duration is much shorter than inverse of the
sterile neutrino production rate $\Gamma_s^{-1} \sim [\sin^2 \theta_s G_F^2 T_\gamma^5]^{-1}$,
only partial thermalization of sterile neutrinos will occur. Interestingly, at the MSW resonance, i.e.,
$\Delta m^2 \simeq  -2EV_{\rm eff}$,
one may get some active-to-sterile neutrino 
(or antineutrino) conversion, depending on the adiabaticity of this resonance. This 
implies that, for $M \gtrsim 10$~MeV, we predict a fractional value of $\Delta N_\text{eff}$ 
at BBN. A study of the detailed dynamics during this epoch is beyond the scope of our present work.

As a final remark, we would like to emphasize that, while Fig.~\ref{fig:Vplot} is
for $E = T_\gamma$, it is important to keep in mind that active neutrinos
follow a thermal distribution. We have checked that even for $E$ different
from $T_\gamma$, the value of $V_\text{eff}$ does not change too much. Therefore,
our conclusions regarding the suppression of sterile neutrino production remain valid
even when the tails of the thermal distribution are taken into account.

\section{Coupling to Dark Matter}
\label{sec:DM}
If a new gauge force of the proposed form exists, it is conceivable that
not only sterile neutrinos, but also DM particles, $\chi$, couple to it.
This of course leads to an additional contribution $2\pi\alpha_\chi(n_{\chi}-n_{\bar\chi})/M^2$ 
to $V_{\rm eff}$, through forward scattering off the net DM density 
(see Appendix A). 
As long as DM is CP-symmetric, we have $n_\chi-n_{\bar\chi} = 0$ and
this extra contribution vanishes. Even for asymmetric DM~\cite{Zurek:2013wia},
we see in Fig.\,\ref{fig:Vplot} (red lines)
that it is usually subleading for $m_\chi\gtrsim 1\,$GeV.
 
The extra gauge interaction of DM does, however, lead to DM self-scattering, 
which has received considerable
attention recently as a way of solving~\cite{Vogelsberger:2012ku, Rocha:2012jg,
Zavala:2012us} the existing disagreement between the observed substructure
of DM in the Milky Way and N-body simulations of galaxy formation. In particular,
self-interacting DM can solve the ``too big to fail'' problem~\cite{BoylanKolchin:2011de,
BoylanKolchin:2011dk}, i.e.,\ the question
why very massive DM subhaloes that are predicted to exist in a Milky Way type
galaxy have not been observed, even though one would expect star formation to
be efficient in them and make them appear as luminous dwarf
galaxies. Similarly, DM self-interactions could be the reason why the Milky Way appears to
have fewer dwarf galaxies than expected from simulations (the ``missing satellites''
problem~\cite{Klypin:1999uc}). Finally, it may be possible to explain why the observed
DM density distribution in Milky Way subhaloes appears to be exhibit a
constant density core~\cite{Moore:1994yx, Flores:1994gz} rather than a steep cusp 
predicted in N-body simulations~\cite{Navarro:1996gj} (``cusp vs.\ core problem'').
While all these problems could well have
different explanations --- for instance the impact of baryonic feedback on
N-body simulations is not yet well understood --- it is intriguing that
the self-scattering cross sections predicted in the scenario discussed here
has exactly the right properties to mitigate these small-scale structure
issues.

In our model, the ``energy transfer cross section'' in the center of
mass frame, $\sigma_T = \int\!d\Omega\,d\sigma/d\Omega (1 - \cos\theta)$, is given in Born approximation by~\cite{Feng:2009hw}
\begin{align}
  \sigma_T \simeq \frac{8\pi\alpha_\chi^2}{m_\chi^2 v_\text{rel}^4}
                  \bigg[ \log(1 + R^2) - \frac{R^2}{1 + R^2} \bigg] \,,
  \label{eq:sigmaT}
\end{align}
with $R \equiv m_\chi v_\text{rel} / M$. Here, $v_\text{rel}$ is the relative
velocity of the two colliding DM particles. It is easy to see that $\sigma_T$ is
velocity independent for $v_\text{rel} \ll M / m_\chi$ and drops
roughly $\propto v_\text{rel}^{-4}$ for larger $v_\text{rel} \gg M / m_\chi$.
This implies that the velocity-averaged cross section per unit DM mass, $\ev{\sigma_T} / m_\chi$,
can be of order 0.1--1~cm$^2$/g in galaxies ($v_\text{rel} \sim \mathcal{O}(100~\text{km/sec})$),
as required to mitigate the small-scale structure problems~\cite{Rocha:2012jg,
Zavala:2012us}, while remaining well below this value in galaxy clusters
($v_\text{rel} \sim \mathcal{O}(1000~\text{km/sec})$), from which the most robust
constraints are obtained\,\cite{Fox:2009in}. 
The cross section given in eq.\,(\ref{eq:sigmaT}) becomes inaccurate in the limit $\alpha_\chi m_\chi/M>1$, and one needs to take nonperturbative/resonant effects into account. In computing $\ev{\sigma_T}$, we take the analytical expressions for $\sigma_T$ for symmetric DM, as summarized in~\cite{Tulin:2013teo}, and convolve with a DM velocity distribution, that we take to be of Maxwell-Boltzmann form, with velocity dispersion $v_{\rm rel}$.

\begin{figure}[!t]
\vspace{-0.1cm}
  \begin{center}
    \includegraphics[width=0.95\columnwidth]{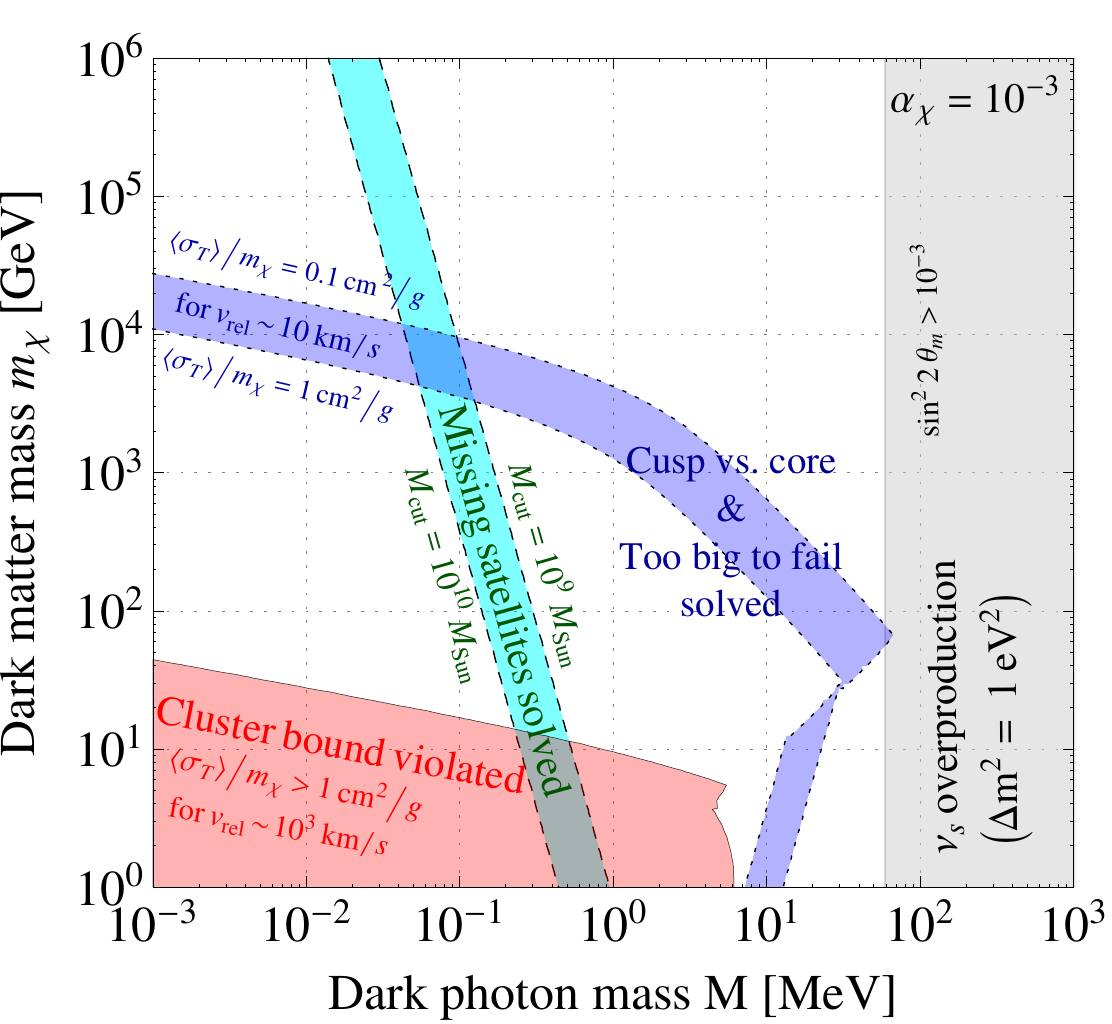}\\[5ex]
  \end{center}
    \vspace{-0.9cm}
  \caption{Constraints on DM self-interactions from the requirements that
    the self-interaction in galaxy clusters is small, i.e., 
    $\langle\sigma_T\rangle / m_\chi \lesssim 1$~cm$^2$/g,  
    and that production
    of 1~eV sterile neutrinos is suppressed, i.e., $\sin^2 2\theta_m \lesssim 10^{-3}$ at 
    $T_\gamma=1\,$MeV. We also show the
    favored parameter region for mitigating the cusp vs. core and too big 
    to fail problems, i.e., $\langle\sigma_T\rangle / m_\chi = 0.1-1$~cm$^2$/g in 
    dwarf galaxies, and solving the 
    missing satellites problem ($M_{\rm cut}= 10^{9-10}\,M_{\rm Sun}$). The kink in the $\sigma_T$ contours is from an approximate treatment of the regime between the Born and classical limits.}
  \label{fig:DM-self-int}
  \vspace{-0.5cm}  
\end{figure}

As for the missing satellites problem, it was shown in~\cite{Boehm:2000gq, Bringmann:2006mu,
Aarssen:2012fx, Shoemaker:2013tda} that DM--neutrino scattering can decrease the temperature of kinetic decoupling of DM, $T_{\rm kd}$, which can increase the cut-off in the structure power spectrum, $M_\text{cut}\propto T_{\rm kd}^{-3}$, to the scales of the dwarf galaxies. $T_{\rm kd}$ is determined by equating the DM momentum relaxation rate $\sim (T_s/ m_\chi) n_s\sigma_{\chi s}$ with the Hubble expansion rate. Here, $n_s \sim T_s^3$ is the sterile neutrino number density, and $\sigma_{\chi s} \sim T_s^2/M^4$ is the DM--sterile neutrino scattering cross section.
Quantitatively~\cite{Aarssen:2012fx},
\begin{align}
\frac{M_\text{cut}}{M_{\rm Sun}}
\simeq3.2\times10^{13}\,\alpha_{_\chi}^{\tfrac{3}{2}}\bigg(\frac{T_s}{T_\gamma}\bigg)^{\tfrac{9}{2}}_{\rm kd}
\bigg(\frac{\rm TeV}{m_\chi}\bigg)^{\tfrac{3}{4}}\bigg(\frac{\rm MeV}{M}\bigg)^{3}\,.
\label{eq:Mcut}
\end{align}
In previous literature, the exponent of $T_s / T_\gamma$
in eq.\,(\ref{eq:Mcut}) is sometimes incorrectly given as
{3/2}~\cite{Bringmann:Private}.
We find the cut-off can be raised to $M_\text{cut} = 10^9-10^{10}\,M_{\rm
Sun}$, as required to solve the missing satellites
problem. The number of sterile neutrino generations $N_s$,
assumed to be 1 here, only weakly impacts the result as $M_{\rm cut}\propto N_s^{3/4}$. 
Note that in contrast to Ref.\,\cite{Aarssen:2012fx}, we obtain a small ${T_s}/{T_\gamma}$, 
from decays of heavy Standard Model particles after the decoupling of the 
sterile sector.
 
In Fig.~\ref{fig:DM-self-int}, we show the region of parameter space favored
by these considerations (see also Appendix B). We see that it is possible to 
simultaneously mitigate the 
cusp vs.\ core problem, too big to fail problem, as well as the missing satellites problem, 
while remaining consistent with the cluster constraint and simultaneously suppressing 
sterile neutrino production to evade BBN and CMB constraints. The potentially interesting 
solution to all the enduring problems with small-scale structures was first shown in a scenario with active neutrinos~\cite{Aarssen:2012fx}, which has since been constrained using 
laboratory data, BBN, and large-scale structure~\cite{Laha:2013xua, Ahlgren:2013wba, Cyr-Racine:2013fsa}. A qualitative extension to sterile neutrinos was suggested therein, and we see here that such a scenario may be realized with no conflict with cosmology.

The DM relic abundance may be produced by Sommerfeld-enhanced annihilations of
DM into $A'$ pairs that decay to sterile neutrinos, or alternatively through an asymmetry. However, unlike
in~\cite{Aarssen:2012fx}, we do not use separate couplings of DM and $\nu$ to
do this, so this should identify the preferred value for DM mass in the range
$m_\chi \sim1-100\,$TeV. As long as DM chemical freeze-out happens well above
$T_\gamma\sim\,$GeV and the sterile neutrinos have time to rethermalize
with ordinary neutrinos (and photons) via high-scale interactions,
our scenario remains unaltered by DM annihilation.

\section{Discussion and Summary}
\label{sec:discussion}
We now discuss the possible origin of a new gauge force in the sterile
neutrino sector, and on further phenomenological consequences (see also
\cite{Bringmann:2013vra}). In~\cite{Pospelov:2011ha},
Pospelov has proposed a model with sterile neutrinos charged under gauged baryon number.
He has argued that the model is consistent with low energy constraints, in particular
the one from $K \to \pi \pi\nu\nu$, even for $\kappa^2 \sin\theta / M^2 \sim 1000 G_F$.
This is precisely the parameter region in which sterile neutrino production
in the early Universe is suppressed, as we have demonstrated above.
In~\cite{Pospelov:2011ha, Harnik:2012ni, Pospelov:2012gm}, the phenomenological
consequences of this model have been investigated, and it has been shown that
strong anomalous scattering of solar neutrinos in DM detectors
is expected.  As an alternative to gauged baryon number, sterile neutrinos could
also be charged under a gauge force that mixes kinetically with the
photon~\cite{Harnik:2012ni}. In this case, $M \gtrsim 10$~MeV is preferred
unless the coupling constants are extremely tiny. Once again in this model
interesting solar neutrino signals in DM detectors can occur. Finally, while
we have focused here on new gauge interactions, it is also conceivable that the
new interaction is instead mediated by a scalar~\cite{Babu:1991at, Enqvist:1992ux}. However, in this case $\sigma_{\chi s}\propto m_{\nu_s}^2$, which is too small and the missing satellite problem cannot be solved.

In summary, we have shown that eV-scale sterile neutrinos can be consistent
with cosmological data from BBN, CMB, and large-scale structure if we allow
them to be charged under a new gauge interaction mediated by a MeV-scale boson.
In this case, sterile neutrino production in the early Universe is suppressed
due to the thermal MSW potential generated by the mediator and by sterile
neutrinos themselves. Our proposed scenario leads to a small fractional number
of extra relativistic degrees of freedom in the early Universe, which may be
experimentally testable in the future. If the considered boson also couples
to DM, it could simultaneously explain observed departures of small-scale
structures from the predictions of cold DM simulations.

\section*{Acknowledgments}
We are grateful to Torsten Bringmann, Xiaoyong Chu, Maxim Pospelov, and 
Georg Raffelt for useful discussions.
JK would like to thank the Aspen Center for Physics, funded by the US National
Science Foundation under grant No.~1066293, for kind hospitality and support
during part of this work. 
We acknowledge the use of the
FeynCalc~\cite{Mertig:1990an} and JaxoDraw~\cite{Binosi:2008ig} packages.

\appendix

\section{Appendix A: Thermal Corrections to Self-Energy}
\label{sec:calculation}

Here, we derive the dispersion relation for sterile neutrinos coupled to a $U(1)_\chi$
gauge force in the regime of nonzero temperature and density. Our approach closely
follows~\cite{Weldon:1982bn,Notzold:1987ik, Enqvist:1990ad,Quimbay:1995jn}.

From considerations of Lorentz invariance, the sterile neutrino self energy
at one-loop can be expressed as
\begin{align}
  \Sigma(k) = (m - a\slashed{k} - b\slashed{u}) P_L \,.
\end{align}
Here, $P_L = (1 - \gamma^5)/2$ is a chirality projector, $m$ is the sterile
neutrino mass, $p$ is its 4-momentum and
$u$ is the 4-momentum of the heat bath. We work in the rest frame of the heat
bath, so we take $u = (1, 0, 0, 0)$.

This thermal self-energy modifies the dispersion relation to
\begin{align}
  \det(\slashed{k} - \Sigma(k)) = 0 \,,
  \label{eq:dispersion1}
\end{align}
which, in the ultrarelativistic regime, $k^0\approx|\vec{k}|$, gives
\begin{align}
k^0=|\vec{k}|+\frac{m^2}{2|\vec{k}|}-b\,
\end{align}
to linear order in the coefficients $a$ and $b$.
Note that the usual dispersion relation for an ultrarelativistic neutrino, 
$k^0=|\vec{k}|+\frac{m^2}{2|\vec{k}|}$, is modified by an effective potential
\begin{align}
  V_\text{eff} \equiv -b \,.
  \label{eq:Veff}
\end{align}
The coefficient $b$ can then be obtained according to the relation
\begin{align}
  b &= \frac{1}{2 \vec{k}^2} \big[ [ (k^0)^2 - \vec{k}^2 ] \tr\,\slashed{u} \Sigma(k) 
                                 - k^0 \tr\,\slashed{k} \Sigma(k) \big] \,.
\end{align}
So, the remaining job is to calculate $\Sigma(k)$. 

\begin{figure}[!h]
  \begin{center}
  \vspace{-0.1cm}
    \includegraphics[width=0.65\columnwidth]{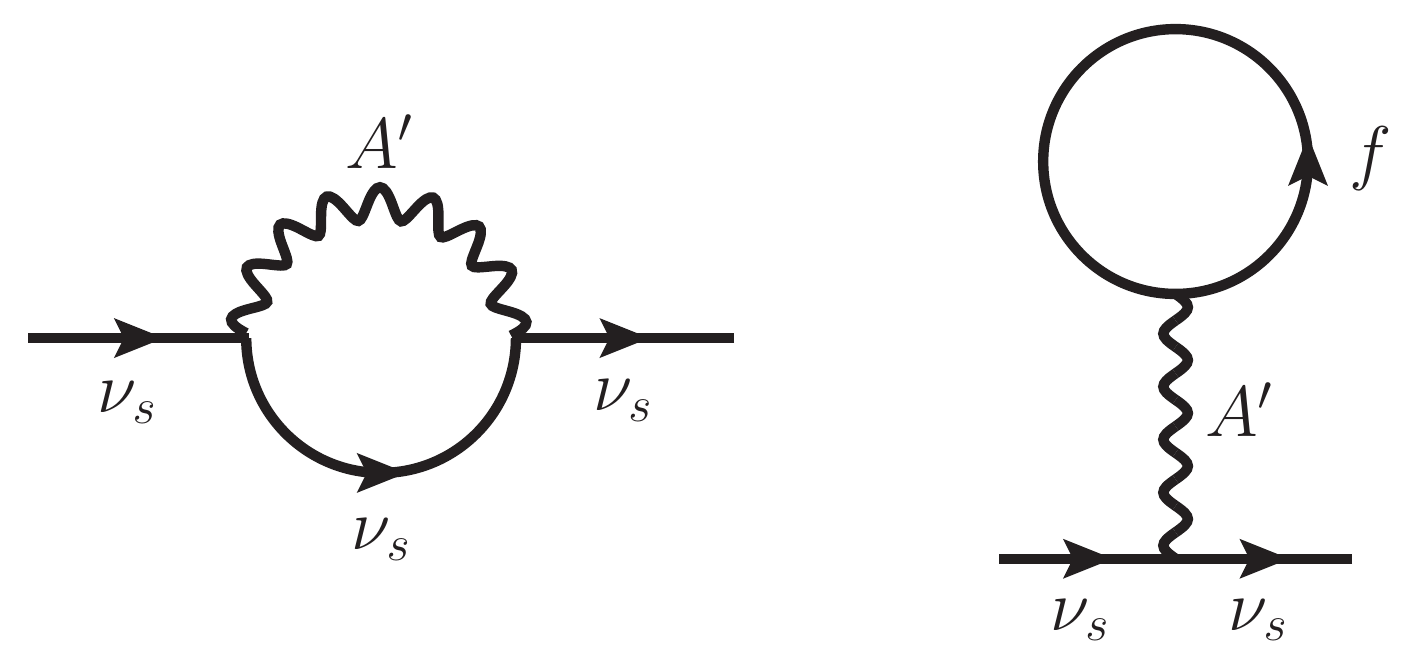}
  \end{center}
  \vspace{-0.0cm}
  \caption{Bubble and tadpole contributions to the sterile neutrino
  self-energy, which create an effective ``matter'' potential.}
    \label{fig:VaFDs}
\end{figure}

We assume a Lagrangian ${\cal L}_{\rm int}=e_\chi \bar{f} \gamma^\mu P_L f A'_\mu$, 
where $e_\chi$ is the $U(1)_\chi$ gauge coupling. At lowest order,
$\Sigma(k)$ receives contributions from the bubble and tadpole
diagrams shown in Fig.~\ref{fig:VaFDs}. In the real time formalism, these
diagrams are calculated using the thermal propagators for the fermion,
\begin{align}
  S(p) &= (\slashed{p}+m) \bigg[ \frac{1}{p^2-m^2} + i\Gamma_f(p) \bigg] \,, \\
\intertext{and the gauge boson (in Feynman gauge)}
  D^{\mu\nu}(p) &= -g^{\mu\nu} \bigg[\frac{1}{p^2-M^2} + i\Gamma_b(p) \bigg]\,.
\end{align}
The thermal parts are given by
\begin{align}
  \Gamma_f(p) &= 2\pi \delta(p^2 - m^2) \eta_f(p) \,, \\
  \Gamma_b(p) &= 2\pi \delta(p^2 - M^2) \eta_b(p) \,,
\end{align}
respectively, with the distribution functions
\begin{align}
  \eta_f(p) &= [e^{|p \cdot u| / T_s} + 1]^{-1} \,, \\
  \eta_b(p) &= [e^{|p \cdot u| / T_s} - 1]^{-1} \,.
\end{align}
The form of $S(p)$ and $D^{\mu\nu}(p)$ can be understood from
the fact that at finite temperature and density, there are not only 
virtual $\nu_s$ and $A'$ in the medium, but also real particles that have 
been thermally excited.

The diagrams in Fig.~\ref{fig:VaFDs} are given by
\begin{align}
  \Sigma_\text{bubble}(k)  &= -i e_\chi^2 \!\int\!\!\frac{d^4 p}{(2\pi)^4} \gamma^\mu \, P_L \,iS(p+k)
                              \, \gamma^\nu \, iD_{\mu\nu}(p)\,, \\
  \Sigma_\text{tadpole}(k) &=  i e_\chi^2 \gamma^\mu \, P_L \, iD_{\mu\nu}(0) \!\int\!\!\frac{d^4 p}{(2\pi)^4} \,
                              \tr\bigg[ \gamma^\nu \,P_L \,iS(p)\bigg]\,.
\end{align}
Since we are interested in the leading
thermal corrections, we evaluate only terms proportional to one
power of $\Gamma_f$ or $\Gamma_b$.  Note that diagrams involving ghosts
do not contribute in the massless sterile neutrino limit.

The leading thermal contributions to the bubble diagram are
\begin{multline}
  e_\chi^2 \int\!\!\frac{d^4 p}{(2\pi)^4} \gamma^\mu (\slashed{k} + \slashed{p}) \gamma^\nu P_L \\
    \times (- g_{\mu\nu})
    \bigg[ \frac{i \Gamma_f(k+p)}{p^2 - M^2} - \frac{i \Gamma_b(p)}{(k+p)^2 - m^2} \bigg] \,.
\end{multline}
We evaluate this expression by first using the $\delta$-functions in $\Gamma_f$
and $\Gamma_b$ to carry out the $p^0$ integral. The remaining 3-momentum integral can be
evaluated in spherical coordinates, with the $z$-axis defined by the direction of $\vec{k}$.
In this coordinate system, the integral over the azimuthal angle is trivial, and the
second angular integral can be evaluated. We have checked that at this stage our
results agree with those of~\cite{Quimbay:1995jn}.

The remaining integral over $|\vec{p}|$ can be carried out numerically, but we derive 
analytical approximations for the two important limiting cases.
In the limit of small temperatures, $|\vec{k}|,\ T_s\ll M$, 
we expand to leading order in $|\vec{p}|/M$, and obtain
\begin{eqnarray}
  b &=&  \frac{7 e_\chi^2 |\vec{k}|}{6\pi^2 M^4}\int_{0}^{\infty} d|\vec{p}|\,|\vec{p}|^3\left(\eta_f + \eta_{\bar{f}}\right)\,\\ 
    &=& \frac{ 7 e_\chi^2 |\vec{k}| \pi^2 T_s^4}{45 M^4} \,. 
  \label{b-largeMA}
\end{eqnarray}
In the opposite limit of high temperature, $|\vec{k}|,\ T_s \gg M$, we can drop subleading logarithmic 
terms in $|\vec{p}|$, and the linear term gives
\begin{eqnarray}
b &=&  -\frac{e_\chi^2}{4\pi^2 |\vec{k}|}\int_{0}^{\infty} d|\vec{p}|\,|\vec{p}| \left(\eta_f + \eta_{\bar{f}} + 2 \eta_b\right)\,\\
  &=&-\frac{e_\chi^2 T_s^2}{8 |\vec{k}|}\,.
\end{eqnarray}
Note that the potential $|b|\ll |\vec{k}|$, thus the neutrinos are still 
ultrarelativistic, and we can replace $|\vec{k}|\approx k^0 = E$, inside the potential. 

In terms of the $U(1)_\chi$ fine-structure constant, $\alpha_\chi\equiv e_\chi^2/(4\pi)$, we thus arrive at,
\begin{align}
  V_\text{eff}^{\rm bubble} \simeq \left\lbrace 
  \begin{array}{lcl}
  -\dfrac{28 \pi^3 \alpha_\chi E T_s^4}{45 M^4}	&\quad& {\rm for}~T_s,E\ll M\\[4ex]
  +\dfrac{\pi \alpha_\chi T_s^2}{2 E}				&\quad&  {\rm for}~T_s,E\gg M
\end{array}
\right.\,
\end{align}
which is the result used in the main text.

Similarly, calculating the tadpole diagram gives
\begin{align}
V_{\rm eff}^{\rm tadpole}\simeq\frac{2\pi\alpha_\chi}{M^2}(n_{f}-n_{\bar{f}})\,,
\end{align}
in terms of terms of the number density of background fermions. It is straightforward to 
see that $\Sigma_\text{tadpole}(k)$ vanishes when there is no fermion asymmetry.
In this work, we have assumed that $\nu_s$ does not have an asymmetry, but instead consider the 
possibility that $A'$ couples to asymmetric DM $\chi$, with a 
net number density, $n_{\chi}-n_{\bar{\chi}}$, which can provide this potential.

\section{Appendix B: Exploration of the parameter space}
\label{sec:mchi}
\begin{figure}[!t]
  \begin{center}
    \includegraphics[width=0.85\columnwidth]{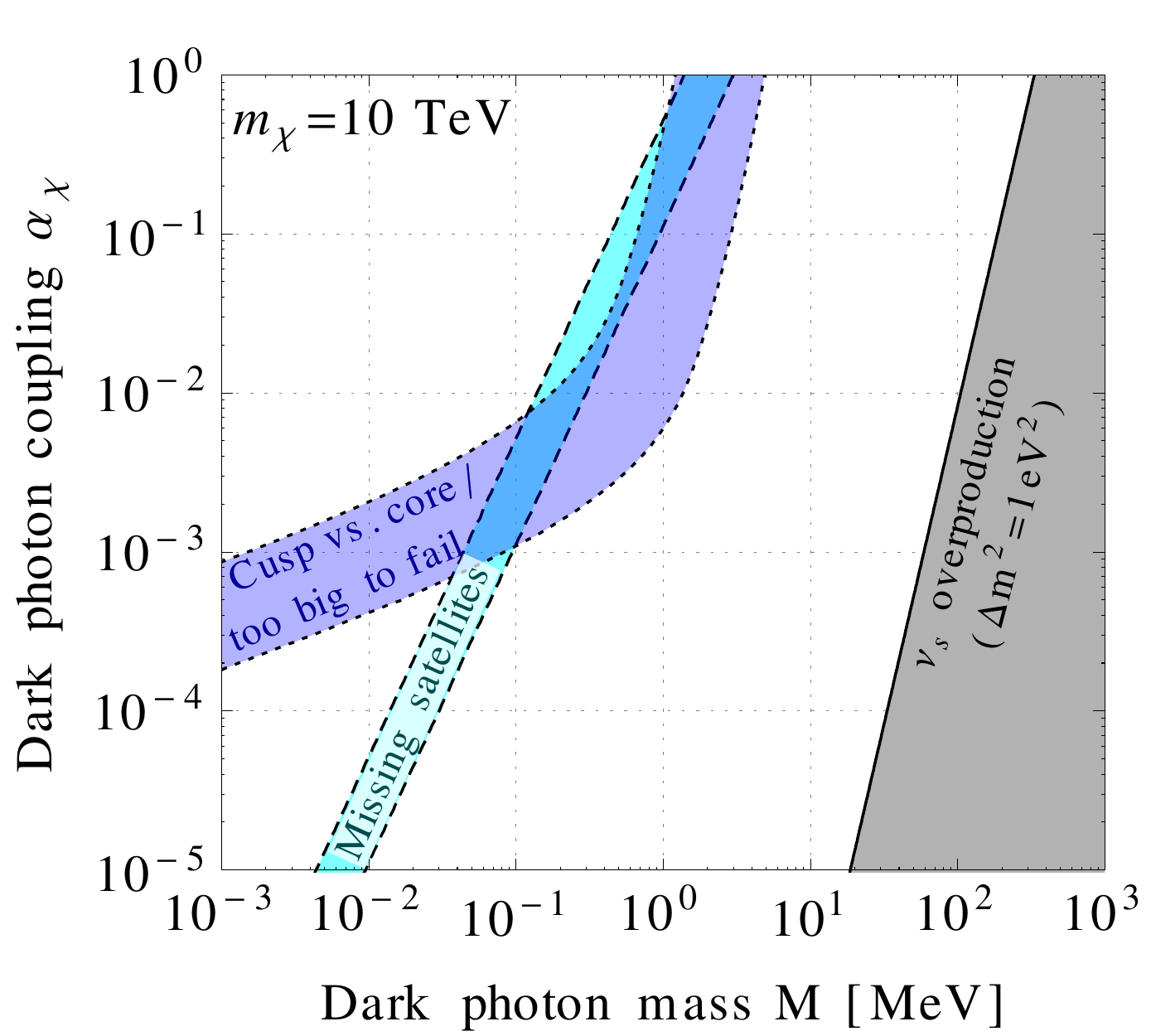}\\[2ex]
    \includegraphics[width=0.85\columnwidth]{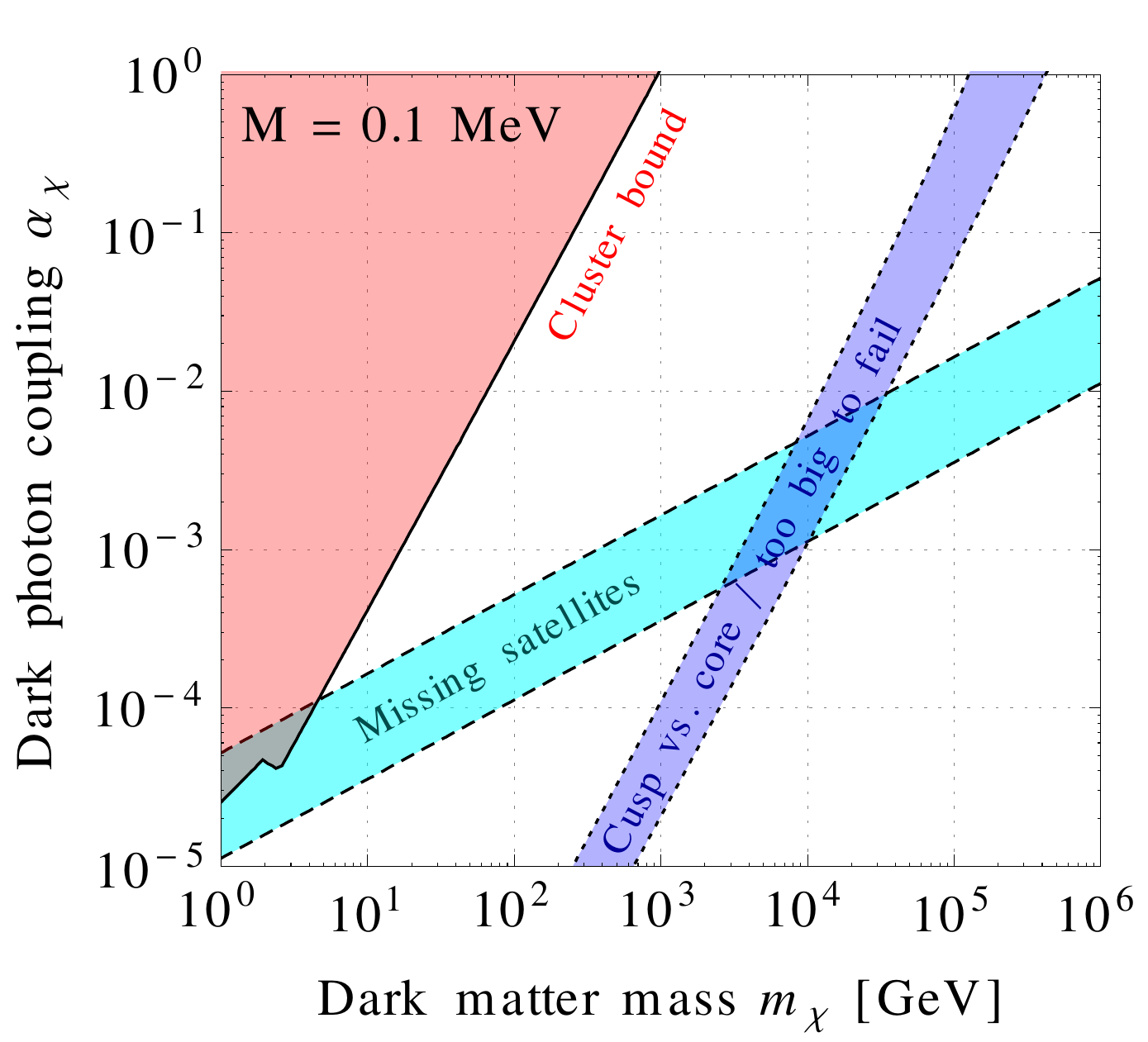}
  \end{center}
    \vspace{-0.5cm}
  \caption{In analogy to Fig.~\ref{fig:DM-self-int}, these plots show the dependence
    of DM self-scattering constraints on the DM coupling for a fixed DM mass 
    $m_\chi=10\,$TeV (top panel) and fixed gauge boson mass $M=0.3\,$MeV (bottom panel).}
  \label{fig:DM-self-int-2}
\end{figure}

In Fig.\,\ref{fig:DM-self-int-2} we show that the DM results, shown in the main text, 
are valid over a reasonable range of values for the coupling $\alpha_\chi$. We find that the favorable 
value for the coupling increases with larger $m_\chi$. The mediator boson mass remains in the 
MeV range. Note that, in the bottom panel of Fig.\,\,\ref{fig:DM-self-int-2}, the neutrino oscillation 
constraints are below the range of the figure.

\bibliographystyle{apsrev}
\bibliography{./Hidden-Sterile}

\end{document}